*Invited Review*

# Phonons and Thermal Transport in Graphene and Graphene-Based Materials

**Denis L. Nika[†‖] and Alexander A. Balandin[†*]**

[†]Phonon Optimized Engineered Materials (POEM) Center and Nano-Device Laboratory (NDL), Department of Electrical and Computer Engineering, University of California – Riverside, Riverside, California 92521 USA

[‖]E. Pokatilov Laboratory of Physics and Engineering of Nanomaterials, Department of Physics and Engineering, Moldova State University, Chisinau MD-2009, Republic of Moldova

## Abstract

A discovery of the unusual thermal properties of graphene stimulated experimental, theoretical and computational research directed at understanding phonon transport and thermal conduction in two-dimensional material systems. We provide a critical review of recent results in the graphene thermal field focusing on phonon dispersion, specific heat, thermal conductivity, and comparison of different models and computational approaches. The correlation between the phonon spectrum in graphene-based materials and the heat conduction properties is analyzed in details. The effects of the atomic plane rotations in bilayer graphene, isotope engineering, and relative contributions of different phonon dispersion branches are discussed. For readers' convenience, the summaries of main experimental and theoretical results on thermal conductivity as well as phonon mode contributions to thermal transport are provided in the form of comprehensive annotated tables.

---
[*] Corresponding author (AAB): balandin@ece.ucr.edu

## I. Introduction

Monoatomic sheet of *sp²* hybridized carbon atoms – graphene – demonstrates unique electrical [1-3], thermal [4-6], optical [7-8] and current fluctuation [9-11] properties owing to its quasi two-dimensional (2D) electron and phonon transport. The ultra-high thermal conductivity of graphene is beneficial for its proposed electronic applications, and it serves as a foundation for numerous possible thermal management applications, e.g. as heat spreaders for transistors and light emitting diodes, and fillers in thermal interface materials for electronic chips [12-19]. There have been several review papers devoted to thermal transport in graphene and graphene-based materials [6, 18, 20-25]. However, the *graphene thermal field* is still in the period of explosive growth. Many new theoretical and experimental results have been reported in the past few years. A number of issues are still awaiting their conclusive resolution. Some new fundamental science questions have been asked. The discussion of the most recent results in the overall context of graphene thermal field is required. In this paper, we review theoretical models for phonons and thermal transport in graphene and graphene nanoribbons (GNRs), describe different experimental techniques for measuring phonon energies and thermal conductivity, and discuss the relative phonon branches contribution to thermal conductivity.

## II. Phonons in graphene and graphene nanoribbons

Single-layer graphene (SLG) possesses the honeycomb crystal lattice with two basis vectors $\vec{a}_1 = a(3,\sqrt{3})/2,$ and $\vec{a}_2 = a(3,-\sqrt{3})/2$, where $a = 0.142$ nm is the distance between two nearest carbon atoms (see figure 1) [26]. The rhombic unit cell of SLG, shown as a dashed region in figure 1, contains two carbon atoms from different Bravais sublattices. In figure 1, the atoms from the first sublattice are marked as "white", while atoms from the second one – as "black" atoms. There are six phonon branches in SLG: in-plane longitudinal acoustic (LA) and optic (LO), in-plane transversal acoustic (TA) and optic (TO) and out-of-plane acoustic (ZA) and optic (ZO). The displacement vector $\vec{U}$ of the in-plane phonon modes has only the in-plane components, i.e. $\vec{U} = (U_x, U_y, 0),$ while $\vec{U}$ of the out-of-plane modes is directed along Z-axis: $\vec{U} = (0,0,U_z)$. Here we assume that axis *X* and *Y* of Cartesian coordinate system are in the plane of graphene layer while axis *Z* is perpendicular to it.

<Figure 1>

The phonon energies are the key parameters for understanding the phonon processes in graphene-based materials. Therefore, significant efforts have been directed at developing various theoretical models for description of phonons in SLG, few-layer graphene (FLG) and graphene nanoribbons [26-64]. These models can be divided into three groups: (i) dynamic models (DMs) of the lattice vibrations [26-46], (ii) elastic continuum models (ECMs) [47-54] and (iii) *ab initio* density functional theory (DFT) models with the local density approximation (LDA) or generalized gradient approximation (GGA) [43, 55-64]. The first group includes the fourth- and fifth-neighbor force-constant models [27, 28, 31-33], valence-force field [26, 29, 30, 34, 35] and Born-von Karman models [36-38], as well as models employing different potentials of the interatomic interaction: Tersoff, Brenner, Lennard-Jones or reactive empirical-bond-order potentials [39-46]. In the framework of DMs, the phonon energies are calculated by solving the system of equations of motion for atoms:

$$\omega^2 u_\alpha^k(\vec{q}) = \sum_{k',\beta} D_{\alpha\beta}^{kk'} u_\beta^{k'}(\vec{q}); \quad \alpha, \beta = x, y, z,$$
$$D_{\alpha\beta}^{kk'} = \frac{1}{m_C} \sum_{n'} \Phi_{\alpha\beta}(k,n;k',n') \exp(\mathbf{i}\vec{q}(\vec{a}_n - \vec{a}_{n'})),$$
(1)

where $u_\alpha^k [u_{\alpha'}^{k'}]$ is a component of the displacement vector for an atom $k[k']$ from the unit cell $n[n']$, $\vec{a}_n [\vec{a}_{n'}]$ is the lattice vector of a unit cell $n[n']$, $\omega$ is the phonon frequency, $\vec{q}$ is the phonon wave vector, $\Phi_{\alpha\beta}(k,n;k',n')$ is the tensor of the second-order interatomic-force constants (IFCs), $m_C$ is a mass of a carbon atom and $k=1,2$. The summation in Eq. 1 is performed over all unit cells $n'$ and over all atoms from the unit cell $k'=1, 2$. The interatomic force-constant tensor strongly depends on the model used, and on the types of the interatomic interactions considered in the model. Usually DMs are characterized by a set of the fitting parameters, which are determined from comparison with the experimental data. If a set of parameters is chosen in an optimal way the dynamic models of the lattice vibrations describe the phonon frequencies with high accuracy. An important difference among DMs is the number of the fitting parameters used. The number of such parameters reported to date varied from 5 [27] to 23 [32].

The phonon dispersion in SLG and AB-stacked bi-layer graphene (AB-BLG) is shown in figure 2 along $\Gamma$-$M$ direction of the Brillouin zone. These dispersion relations were calculated using the Born-von Karman approach for the intralayer carbon-carbon interactions [36] and the spherically symmetric interatomic potentials for the interlayer interactions [38]. The red triangles show the experimental frequencies from Ref. [28]. At small $q$ in the vicinity of $\Gamma$-point, the *LA* and *TA* bracnhes are almost linear with $q$: $\omega_{LA,TA} \sim q$ while the ZA branch demonstrates quadratic dependence $\omega_{ZA}(q) \sim q^2$.

<Figure 2>

In the framework of the elastic continuum models, the few-layer graphene is approximated as a stack of equally spaced and interacting elastic sheets. The system of equations of motion for the elastic vibrations in each sheet is given by [48]:

$$D\Delta^2 w_i + \rho h \frac{\partial^2 w_i}{\partial t^2} = q_i, \quad i=1,2,\ldots,N \qquad (2)$$

where $N$ is the number of sheets, $D$ is the bending stiffness of each sheet, $\rho$ is the mass density, $h$ is the sheet thickness, $q_i$ is the pressure applied to an atomic sheet $i$ due to the interlayer van der Waals (vdW) interactions, $w_i$ is the deflection of $i$-th sheet and $\Delta = \partial^2/\partial x^2 + \partial^2/\partial y^2$. For infinitesimal vibrations, the pressure due to the vdW interactions can be assumed in the following form: $q_i = w_i \sum_{j=1}^{N} c_{ij} - \sum_{j=1}^{N} w_j c_{ij}$, where $c_{ij}$ are the vdW interaction coefficients [48]. Solving Eq. (2) by using two-dimensional propagating waves, the following equations for phonon frequencies were derived in Ref. [48]:

$$\left( D\left[ \left(\frac{\pi m}{L_x}\right)^2 + \left(\frac{\pi n}{L_y}\right)^2 \right]^2 - \sum_{j=1}^{N} c_{ij} - \rho h \omega^2 \right) u_i + \sum_{j=1}^{N} c_{ij} u_j = 0, \quad i=1,2,\ldots,N, \qquad (3)$$

where $L_x$ is the sheet length, $L_y$ is the sheet width, $m$ and $n$ are the half wave numbers in the sheet plane.

Employing the semi-continuum model from Ref. [65], Nishira and Ivata [47] derived the analytical expressions for the phonon frequencies in graphite:

$$\omega_{LA}^2 = v_l^2(q_x^2 + q_y^2) + \frac{4\zeta}{c^2}\sin^2\left(\frac{cq_z}{2}\right),$$

$$\omega_{TA}^2 = v_t^2(q_x^2 + q_y^2) + \frac{4\zeta}{c^2}\sin^2\left(\frac{cq_z}{2}\right), \quad (4)$$

$$\omega_{ZA}^2 = b^2(q_x^2 + q_y^2)^2 + 4\mu^2\sin^2\left(\frac{cq_z}{2}\right) + \zeta(q_x^2 + q_y^2).$$

In equation (4) $v_l/v_t$ is the longitudinal/transverse in-plane velocity, correspondingly, $c$ is the interlayer spacing, $b$ is the bending elastic parameter, $\zeta = c_{44}/\rho$ and $\mu = \sqrt{c_{33}/(\rho c^2)}$. In the case of single-layer graphene, $q_z = 0$ and the phonon frequencies are given by:

$$\omega_{LA}^{SLG} = v_l\sqrt{(q_x^2 + q_y^2)} = v_l q,$$

$$\omega_{TA}^{SLG} = v_t\sqrt{(q_x^2 + q_y^2)} = v_t q, \quad (5)$$

$$\omega_{ZA}^{SLG} = \sqrt{b^2(q_x^2 + q_y^2)^2 + \zeta(q_x^2 + q_y^2)}.$$

<Figure 3>

In figure 3 we show the dispersion for the *LA*, *TA* and *ZA* phonon branches along the *Γ-M* direction of BZ, calculated from Eq. (5). The parameters $v_l = 21.6$ km/s, $v_t = 14.0$ km/s, $\zeta = 1.88 \times 10^{10}$ cm²/s² and $b = 3.13 \times 10^{-3}$ cm²/s were taken from Ref. [47]. In the semi-continuum model $\omega_{LA}^{SLG}, \omega_{TA}^{SLG} \sim q$ while $\omega_{ZA}^{SLG} \sim q^2$ over the entire Brillouin zone resulting in an overestimation of the *LA/TA* phonon frequencies for $q > 8$ nm$^{-1}$ and underestimation of ZA phonon frequencies for $q > 4$ nm$^{-1}$ in comparison with both experimental and BvK model results (see figure 2). The deviation of the ZA phonon branch from the parabolicity (see figure 2) strongly affects the phonon density of states and specific heat [38].

The *ab initio* DFT-based models are a powerful tool for investigation of phonon processes in graphene materials. Nevertheless, employment of these models requires careful checking of the obtained results. Possible numerical inaccuracies in calculating the *ab initio* IFCs may strongly influence the phonon dispersions, and even lead to deviation from the $q^2$ dependence of the ZA branch at small *q* [66]. To resolve this issue, in Ref. [67] the acoustic sum rules were numerically reinforced.

Although the general trends in the phonon dispersion in graphene are well described by the majority of theoretical models, the fine behavior of phonon branches as well as phonon energies in high symmetry points of Brillouin zone very strongly vary from one model to another [20]. This discrepancy is one of the reasons for different values of the phonon scattering rates and thermal conductivity predicted by different theoretical approaches [20]. An increase in the number of graphene atomic layers results in the larger unit cell volume and the number of atoms in the unit cell. The unit cell of the *n*-layer Bernal stacked graphene contains $2n$ atoms, therefore $6n$ phonon branches appear in the energy spectrum of *n*-layer graphene sample.

<Figure 4>

The evolution of the phonon energy spectra near center of the BZ when one goes from SLG to AB-bilayer graphene (AB-BLG) to ABA-trilayer (ABA-TLG) graphene is illustrated in figure 4. Additional phonon branches of AB-BLG and ABA-TLG, indicated in figure 4 as *LA$_2$*, *LA$_3$*, *TA$_2$*, *TA$_3$*, *ZA$_2$* and *ZA$_3$*, are characterized by non-zero frequencies at $\Gamma$ point. These frequencies strongly depend on week interlayer vdW forces, therefore their values vary from one theoretical model to another. Employment of the Lennard-Jones potential for modeling of vdW interlayer interaction in graphite results in lower frequencies of *LA*/*TA* branches, calculated along $\Gamma A$ direction of the BZ, as compared with the experimental data [37]. Authors of Refs. [37-38] proposed to model interlayer interaction by spherically symmetric interatomic potential instead of Lennard-Jones potential. The phonon dispersions in graphite calculated within the model with spherically symmetric interatomic potential were in a good agreement with the experimental curves [37].

The phonon modes in twisted bilayer graphene (TBLG) were theoretically studied in Ref. [36]. When two graphene layers are placed on top of each other they can form a Moire pattern [68-70]. In this case, one layer of carbon atoms is rotated relative to another layer by an arbitrary angle. The size of the unit cell in TBLG is larger than in AB-BLG and depends strongly on the rotational angle $\Theta$. Therefore the number of phonon branches also depends on $\Theta$. In figure 5 we show the phonon dispersions calculated along $\Gamma$-$K$ direction of the BZ in AB-BLG (a) and TBLG with $\Theta=21.8°$ (b). The Brillouin zone of TBLG with $\Theta=21.8°$ is by a factor of seven smaller than the BZ of AB-BLG. Therefore, the *hybrid folded* phonon branches appear in twisted bilayer graphene resulting from mixing of different directions from un-rotated BLG BZ [36]. The twisting also slightly affects the phonon frequencies in TBLG due to modification of the weak vdW interlayer interaction.

<Figure 5>

The peculiarities of the phonon energy spectra in graphene reveal themselves in the phonon density of states (PDOS) and phonon specific heat. The 2D phonon density of states is given by [38]:

$$g(\omega) = \sum_s g_s(\omega); \quad g_s(\omega) = \frac{1}{4\pi^2} \sum_{q_x(s,\omega)} \sum_{q_y(s,\omega,q_x)} \frac{\Delta q_x}{|\upsilon_y(q_x,q_y,s)|}. \quad (6)$$

Here, $s$ numerates phonon branches (polarizations), $g_s(\omega)$ is the polarization-dependent phonon density of states, $q_x$ and $q_y$ are components of the 2D phonon wave vector, $\upsilon_y = \partial\omega/\partial q_y$ is the y-component of the phonon group velocity, $\Delta q_x$ is the interval between two neighboring $q_x$ points. The dependence of PDOS in SLG (solid curves) and AB-BLG (dashed curve) on phonon frequency is presented in figure 6. The contribution of different phonon branches *LA* (green), *TA* (blue), *ZA* (red) and *ZO* (magenta) are also shown. The PDOS peaks at 452, 605 and 638 cm$^{-1}$ correspond to *ZA*, *TA* and *ZO* phonon branches at BZ edge, correspondingly; the *LA* peak at ~ 1192 cm$^{-1}$ is associated with the low-velocity *LA* phonons from different directions near BZ edge; the peak at 889 cm$^{-1}$ is related to *ZO* phonon at $\Gamma$ point and TO and LO phonons at BZ center and BZ edge are responsible for peaks at 1350 and 1585 cm$^{-1}$. The peak at 91 cm$^{-1}$ of AB-BLG PDOS, which is absent in SLG, is related to ZA$_2$ phonons at $\Gamma$ - point.

<Figure 6>

Theoretical studies of the phonon specific heat $c_v$ in SLG were carried out in Refs. [37, 38, 52, 71]. Authors of Refs. [52, 71] used a simple model of phonon dispersions in graphene: parabolic *ZA* dispersion $\omega_{ZA} \sim q^2$ and linear *LA/TA* dispersions $\omega_{LA,TA} \sim q$ over entire Brillouin zone. In this case, the low-temperature specific heat is proportional to $T$ for ZA modes and to $T^2$ for LA/TA modes. The linear dependence of total specific heat on temperature $c_v \sim T$ was reported in Ref. [71] for $T<100$ K. The slight deviation from the linear $T$ dependence due to contribution from LA and TA phonons was obtained in Ref. [52]: $c_v \sim T^{1.1}$. However more rigorous model of phonon specific heat in SLG, which takes into account both anisotropy of phonon dispersions and non-parabolicity of ZA branch for $\omega >100$ cm$^{-1}$, shows that the deviation from linear $T$ dependence is stronger for T>35 K: $c_V \sim T$ for $T \leq 15\,\text{K}$; $c_V \sim T^{1.1}$ for $15\,\text{K} < T \leq 35\,\text{K}$; $c_V \sim T^{1.3}$ for $35\,\text{K} < T \leq 70\,\text{K}$ and $c_V \sim T^{1.6}$ for $75\,\text{K} < T \leq 240\,\text{K}$ [37,38]. In bilayer graphene the low-temperature specific heat scales with $T$ as $c_V \sim T^n$, where $n$=1.3 for AA- or AB-BLG and $n$ = 1.6 for TBLG with Θ=21.8° [38].

<Figure 7>

The temperature dependence of the specific heat in SLG and AB-BLG is depicted in figure 7 (a). The experimental points for graphite from Ref. [47] are also shown by the blue triangles. The difference between the heat capacities in graphite, SLG and BLG decreases with increasing temperature, and for $T$>2500 K all heat capacities approach the classical Dulong-Petit limit $c_v$=24.94 JK$^{-1}$mol$^{-1}$. The dependence of $c_v$ in TBLG on the rotational angle is illustrated in figure 7(b). In this figure we plot a difference between the specific heats in AB-BLG and TBLG $\Delta c_v = c_v(AB-BLG) - c_v(TBLG)$ as a function of temperature for different Θ. The change in specific heat due to twisting is relatively weak for temperatures $T$>20 K. Nevertheless at very small temperature $T$~1 K, the relative difference between specific heat in AB-BLG and TBLG $\eta = \Delta c_v / c_v(AB-BLG) \times 100\%$ achieves 10-15% (see inset to figure 7(b)) due to changes in the frequencies of ZA phonons.

The experimental investigations of phonon modes in graphite, single- and few-layer graphene were carried out using inelastic x-ray scattering [28, 56], inelastic electron tunneling spectroscopy [72] and Raman spectroscopy [73-89]. The earlier Raman studies of SLG revealed three different phonon bands in graphene: *G*-, *D*- and *2D*-band. The *G*-band near 1485 cm$^{-1}$ is the first-order Raman peak associated with scattering of in-plane optical phonons of BZ *Γ*-point. Scattering of two *TO* phonons around *K*-point of BZ gives rise to 2D-band, i.e. the second-order Raman peak in the range 2500 – 2800 cm$^{-1}$. More recent Raman studies observed peaks associated with the shear phonons in few-layer graphene [78-82] and folded phonons in twisted few-layer graphene [69-70, 83-89]. The dependence of the Raman 2D peak spectral position and shape on number of carbon atomic layers is shown in figure 8.

<Figure 8>

## III. Two-dimensional thermal transport in graphene-based materials

In this section we review theoretical and experimental results pertinent to the phonon thermal transport in graphene – based materials, focusing on the most recent reports. These findings are discussed in details and compared with earlier results.

### *III.1. Experimental investigations of thermal transport in graphene and graphene-based materials*

The first experimental measurements of thermal conductivity of graphene were conducted at the University of California – Riverside in 2008 [4-5] using non-contact Raman optothermal method. In this technique the central part of graphene layer suspended over a trench was heated by the laser light, resulting in local temperature rise and corresponding frequency shift of Raman G peak. Measuring this frequency shift allows one to extract the temperature profile of local heated area. The measured room temperature (RT) values of thermal conductivity (TC) were in the range $\kappa$=3000 – 5400 W/mK and exceeded the thermal conductivity of the best bulk thermal conductors: highly – oriented pyrolitic graphite (HOPG) and diamond [6]. In figure 9 we illustrate the linear dependence of Raman G peak shift on the temperature, which is used for extraction of the temperature in the Raman optothermal method.

<Figure 9: Raman shift vs Temperature>

The following independent measurements of thermal conductivity in graphene were performed using different techniques: Raman optothermal, electrical self-heating and T-bridge methods. Using Raman optothermal technique Cai et al. [90] found that thermal conductivity of suspended chemical vapor deposited (CVD) graphene is ~ 2500 W/mK at 350 K and ~ 1400 W/mK at 500 K. Other optothermal studies revealed TC in suspended CVD graphene in the range from 1500 to 5000 W/mK [91]. Faugeras et al. [92] also employed Raman optothermal method for measurements of thermal conductivity of suspended graphene in Corbino membrane geometry and found $\kappa$ ~ 600 W/mK at $T$ ~ 660 K. As in conventional materials the thermal conductivity decreases with temperature owing to increasing phonon – phonon scattering.

The high-temperature thermal transport in suspended exfoliated and CVD graphene was experimentally studied by Dorgan et al. [93] within electrical self-heating method. For this study authors fabricated 15 devices with suspended exfoliated or CVD graphene. Average TC of exfoliated and CVD graphene samples were similar $\kappa$ ~ 310 + 200/-100 Wm$^{-1}$K$^{-1}$ at $T$ = 1000 K. The RT TC was in the range 2000 – 3800 Wm$^{-1}$K$^{-1}$ with the average value of 2500 W/mK, which is in a good agreement with previous experimental results [4-5, 90-91]. Dorgan et al. [93] also found that high-temperature TC in graphene demonstrates steeper decrease with temperate $\kappa$ ~ $T^{1.7}$ than that in graphite. This effect was attributed to the stronger second-order three phonon scattering in graphene. The lower values of thermal conductivity were found in the supported graphene due to coupling of graphene phonon modes to substrate modes and additional phonon scattering on the graphene – substrate interface [94]. The thermal conductivity of graphene encased in other materials also substantially smaller than that in SLG [95]. The reasons for TC drop are similar to the supported graphene case: coupling of graphene phonons to phonons from another material and phonon scattering on interfaces and disorder. The dependence of the thermal conductivity of suspended graphene on the number of atomic layers $n$ was studied in Ref. [34]. It was established that thermal conductivity decreases with increasing of $n$ from 1 to 4 and for $n$=4 approaches the value of TC in HOPG. The polymeric residues, often presenting on graphene surface influence both TC values [96] and TC dependence on the number of atomic layers [97]. The unusual dependence of the intrinsic thermal conductivity of graphene on the number of atomic planes was discussed in details in Ref. [6].

The isotopic scattering is an important parameter, which affects the graphene thermal conductivity. Chen et al. [98] reported on experimental study of isotope effect on thermal properties of graphene, using the Raman optothermal method. The increase of $^{13}$C isotope concentration $N_{isot}$ led to the strong suppression of the thermal conductivity from ~ 2800 W/mK for $N_{isot} = 0.01\%$ to ~ 1600 W/mK for $N_{isot} = 50\%$ at $T$ ~ 380 K (see figure 10).

<Figure 10: Isotope effect>

Li et al. [99] also employed Raman optothermal technique for investigation of thermal transport in twisted bilayer graphene. The authors found that in a wide range of examined temperatures, from 300 K to 750 K, the TC in T-BLG is smaller than both in SLG and AB-BLG (see figure 11). The thermal conductivity of twisted bilayer graphene is by a factor of two smaller than that in SLG and by a factor of ~ 1.35 smaller than that in AB-BLG near the room temperature. The drop of TC was explained by emergence of many additional hybrid folded phonons in T-BLG resulting in more intensive phonon scattering [99].

<figure 11: K in T-BLG>

Experimental studies [100-102] reported on strong dependence of thermal conductivity in graphene and GNRs on the sample size: length or width. Xu et al. [101] carried out measurements of thermal conductivity in suspended CVD single-layer graphene, using the electro-thermal bridge method and observed the logarithmic dependence of TC on the sample length $L$ $\kappa \sim \log(L)$ for examined range of $L$ between 700 nm and 9 µm. Although different thermal resistivity of samples with different $L$ may have affected the reported results, it is interesting to note that the obtained $\kappa \sim \log(L)$ dependence of thermal conductivity was in agreement with earlier theoretical predictions made for graphene [103-105] and pure 2D lattices [106-107].

Bae at al. [100] found that RT TC in GNRs with length $L \approx 260$ nm drops from 230 to 80 Wm$^{-1}$K$^{-1}$ with decrease of GNRs width $W$ from 130 to 45 nm, respectively, due to enhancement of edge roughness scattering. More general, the authors predicted that in GNRs with $L$ and $W$ larger than phonon MFP $\lambda$, the thermal transport is diffusive. In GNRs with $L \sim \lambda$ and $W \gg \lambda$, the transport is quasi-ballistic, while in GNRs with $L \sim \lambda$, $W \sim \lambda$ and $L > W$ the transport is diffusive due to phonon edge roughness scattering. These findings are in a qualitative agreement with theoretical results reported for micrometer graphene ribbons [108].

A more recent study by Chen et al. [102] revealed the opposite effect of decreasing TC with increasing sample width for micrometer-wide graphene ribbons (GR). The authors claimed that TC increases from 205 W/mK in SLG to 2236.26 W/mK in GR with $W$~43-50 µm at room temperature. Several possible reasons for such behavior have been proposed and discussed: excitation of more low-frequency phonon modes with $W$ decrease or change in the phonon – edge localization. However, additional experimental and theoretical works are required to establish the accurate scenario. The validity of Fourier's law for graphene was analyzed in Ref. [109]. Jo et al. concluded [109] that linear dependence of thermal resistance on sample length, measured by Xu et al. [101] does not reveal the failure of Fourier's law. The authors of Ref. [109] also measured the thermal conductivity in suspended exfoliated bi-layer graphene, using electro-thermal micro-bridge method and found TC in the range $(730 – 880) ± 60$ $Wm^{-1}K^{-1}$ at RT.

Another experimental study [110] employed four-wire electrical self-heating method to measure the thermal conductivity in a 169-nm wide and 846-nm long graphene ribbon. The temperature dependence of TC $\kappa \sim T^{2.79}$ was reported for temperature range 80 – 380 K, while $\kappa \sim T^{1.23}$ was revealed for low temperatures. The measured values of TC varied from $(12.7 ± 2.95)$ $Wm^{-1}K^{-1}$ at 80 K to $(932±333)$ $W m^{-1}K^{-1}$ at 380 K. The TC ~ $(349 ± 63)$ $W m^{-1}K^{-1}$ found in this study is substantially lower than that in the large suspended graphene layers but it is in an agreement with TC reported for graphene ribbons [20,33,101,105,108,111]. The deviation of the measured data from the quasi-ballistic transport limit allowed the authors to conclude that in the considered narrow and short GNR the thermal transport is diffusive due to phonon-edge scattering.

The described experimental data confirm that graphene as superior thermal conductor is a promising material for the thermal management applications. However, production of the large high-quality graphene sheets is still a major technological challenge. The research community continues to search for inexpensive graphene-based materials with sufficiently high thermal conductivity. Recently, it was demonstrated that graphene laminate (GL) [112], reduced graphene oxide (rGO) [113] and graphene paper (Gp) [114] annealed at high temperatures possess high in-plane thermal conductivity and may be used for the thermal management as heat spreaders or fillers in the thermal interface materials. The graphene derived materials are multilayered structures of carbon layers with good in-plane interaction between atoms and weak interlayer coupling. The RT TC varies from ~ 60 W/mK for rGO [113] to ~ 40 – 90 W/mK for GL [112] and to ~1400 W/mK for Gp [114]. The thermal conductivity in GL and rGO strongly depends on lattice defects and average size of grains/carbon clusters. High temperature treatment of rGO samples leads to simultaneous increase of in-plane thermal conductivity and decrease of out-of-plane thermal conductivity, resulting in exceptionally strong anisotropy of the thermal conductivity $\kappa_{in-plane}/\kappa_{out-of-plane}$ ~ 675, which is by a factor of ~ 6.7 larger even than in the HOPG [113]. The dependece of thermal conductivity on average cluster/grain size in GL and rGO is illustrated in figure 12.

<Figure 12 (a,b): K in GL and rGO>

The experimental data on the thermal conductivity of graphene and graphene-based materials are summarized in Table 1. The TC values are for RT unless another temperature is indicated.

*III.2. Theoretical models of thermal transport in graphene and graphene nanoribbons*

The unique features of phonon transport in 2D and intensive experimental investigations stimulate theoretical studies in the graphene thermal field. The theoretical models employed for the investigation of heat conduction in graphene and GNRs can be rougly devided into two groups: Boltzmann transport equation (BTE) approach and molecular dynamics (MD) simulations, which include eqilibrium molecular dynamics (EMD) or nonequlibrium molecular dynamics (NEMD). These models have been used in numerous theoretical studies of thermal conductivity in graphene and GNRs, which focused on the thermal conductivity dependence on flake size, defects, isotopes, strain, grain size and amharmonicity of crystal lattice.

The initial BTE-based theoretical investigations of heat conduction in graphene were carried out within the relaxation time approximation (RTA) and long-wave length approximation (LWA) for three-phonon Umklapp scattering rates. We will refere hearafter to this approch as BTE-LWA. In his seminal works [103-104], Klemens concluded that therma transport in graphene sheet is two-dimensional down to zero frequency, and, therefore, the intrinsic thermal conductivity limited by the three-phonon Umklapp scattering with the scattering rate $1/\tau_U \sim \omega^2$ demonstrates logarithmic divergence. He proposed to limit the length of the long wave-length phonons by the average size of graphene sheet *L*. The latter avoids the TC divergence, resulting in the dependence of TC on the extrinsic parameter *L*: $\kappa \sim \log(L)$. Using simple isotropic phonon dispersion $\omega = <v> q$, where $<v>$ = 18.6 km/s and average value of the Gruneisen anharmonicity parameter $\gamma=4$, Klemens calculated the room-temperature $\kappa=4400$ W/mK for graphene sheet with $L = 1$ mm, which is in agreement with the first experimental findings [4-5]. Nika et al. [105] modified Klemens model by using a more general expression for the thermal conductivity, introducing two different average Gruneisen parameters for LA and TA branches and taking into account the difference in the phonon group velocity between them. The strong dependence of TC on *L*, Gruneisen parameters and temperature were predicted. It was found that the increase in *L* from 1 µm to 50 µm enhances the RT TC from 1000 to 8000 W/mK. Despite its simplicity Klemens-like BTE-LWA model describes the size dependence of TC rather accurately and in line with both experimental studies [4-5,100-101] and more rigorous theoretical models developed later [26,40-41,108]. Nevertheless, this model does not take into account 2D specifics of the three-phonon Umklapp scattering, which is crucial for understanding the phonon transport in graphene. More elaborate BTE-LWA calculations of the three-phonon Umklapp scattering rates, which considered all possible three-phonon transitions in graphene, allowed by the energy and momentum conservation, were reported in Ref. [26]. In this case the Umklapp scattering rate was given by:

$$\frac{1}{\tau_U^{(I),(II)}(s,\vec{q})} = \frac{\hbar \gamma_s^2(\vec{q})}{3\pi \rho v_s^2(\vec{q})} \sum_{s's'';b_i} \iint \omega_s(\vec{q}) \omega_{s'}'(\vec{q}') \omega_{s''}''(\vec{q}'') \times \{N_0[\omega_{s'}'(\vec{q}')] \mp \\ \mp N_0[\omega_{s''}''(\vec{q}'')] + \frac{1}{2} \mp \frac{1}{2}\} \times \delta[\omega_s(\vec{q}) \pm \omega_{s'}'(\vec{q}') - \omega_{s''}''(\vec{q}'')] dq_l' dq_\perp', \quad (7)$$

In equation (7) the upper signs correspond to the three-phonon processes of the first type, when a phonon with the wave vector $\vec{q}(\omega)$ absorbs another phonon from the heat flux with the wave vector $\vec{q}'(\omega')$, forming the phonon with wave vector $\vec{q}''(\omega'')$ in one of the nearest Brillouin zones; the lower signs correspond to those of the second type, when phonon $\vec{q}(\omega)$ of the heat flux decay into two phonons with the wave vectors $\vec{q}'(\omega')$ and $\vec{q}''(\omega'')$ in one of the nearest BZ. The integrals for $q_l, q_\perp$ are taken along and perpendicular to the curve segments, correspondingly, where the conditions of the energy and momentum conservation are met [26]. Using this formalism, Nika et al. [26] found a strong dependence of the thermal conductivity on temperature, point-defects, size and edge roughness of the flake. Depending on these parameters the TC values from 2000 to 12000

W/mK were obtained. It was also found that LA and TA phonons are the main heat carriers in graphene. The contribution from ZA phonons was small due to the large negative values of the Gruneisen parameters for the long and medium wave-length ZA phonons, resulting in their strong scattering. The latter together with the small group velocities of ZA phonons led to their smaller contribution to the phonon heat flux as compared to LA and TA phonon modes. However, Eq. (7) was obtained in the long-wave length approximation. Its validity for the medium and short wave-length phonons is limited. It also contains Gruneisen parameter $\gamma_s(\vec{q})$, which depends only on the anharmonic property of the phonon $\vec{q}(\omega)$, and it is averaged over the anharmonic properties of phonons $\vec{q}'(\omega')$ and $\vec{q}''(\omega'')$.

In several followed works, Lindsay et al. [40-41] also employed BTE approach within RTA, but without LWA for the matrix elements of three-phonon scattering. In their approach, Lindsay et al. calculated the third-order interatomic force constants for each phonon mode and found that ZA phonons carry ~ 75 % of heat in graphene. The three-phonon matrix elements, derived beyond the LWA, also imply the special selection rules for ZA phonons scattering: the participation of odd ZA phonons in three-phonon transitions is forbidden [40-41]. These selection rules were not taken into account in the previous studies [26,103-105], resulting in an overestimation of the scattering intensity for ZA phonons and understimation of their contribution to TC. We will refer below the theoretical approach developed by Lindsay et al. [40-41] as BTE-IFC. It is important to note here that both BTE-LWA and BTE-IFC approaches predict strong dependece of the thermal conductivity in graphene on the extrinsic parameters: edge roughness, defects and flake size. The reported values of the thermal conductivities in these approaches are in a good agreement with each other as well as with experimental data [4-5,90-91].

The first theoretical studies of thermal conducitvity in graphene within MD simulations were reported in Refs. [118-120]. Using equilibrium molecular dynamics with the Brenner-type of bond order dependent potential, Che et al. [118] studied RT thermal conductivity of carbon nanotubes (CNTs) and compared them with the calculated graphene thermal conductivity $\kappa$ ~ 1000 W/mK. Osman and Srivastava [119] investigated thermal conductivity of CNTs and graphene using MD with the Tersoff-Brenner potential for C-C interaction. The RT TC of graphene $\kappa$ ~ 1500 W/mK was obtained in their calculations. One should note that in most of the MD calculations the absoliute values of TC are limited by the size of the simulated sample. Berber et al. [120] reported on higher value of RT TC in graphene $\kappa \sim 6600$ W/mK withing combined equilibrium and noneqiulibrium MD with Tersoff interatomic potential. The thermal transport in GNRs was considered in Refs. [121-123] in the framework of EMD and NEMD with Brenner [121], Tersoff [122] and Stillinger-Veber type [123] potentials. It was established that in few-nanometer size GNRs, the thermal conductivity depends on edge chirality and defects of crystal lattice [121]. Zigzag edge GNRs (ZGNRs) demonstrated higher thermal conductivity than arm-chair edge GNRs (AGNRs), while defects suppressed thermal conductivity twisely [121] from $\kappa \sim 1400$ W/mK to 700 W/mK at RT. The length and strain dependence of TC in 20-AGNR and 10-ZGNR was investigated in Ref. [122]. In the considered range of the length from 10 nm to 60 nm, the TC increased with $L$ as $\kappa \sim L^n$, where $n = 0.47$ for 20-AGNR and $n=0.35$ for 10-ZGNR [122]. The 15% - strain decreased TC by a factor of $\sim 4.5$ for 10-ZGNR and by a factor of $\sim 2$ for 20-AGNR. The strong thermal rectification effect in GNRs with different shapes was also elucidated [121,123]. Readers interested in more detailed description of theoretical results on thermal transport in graphene and GNRs, reported between 2009 and 2012 are reffered to different reviews [6, 20-24]. Below we focus on the review of more recent theoretical results, their discussion and comparison with earlier fundings.

The thermal transport in bicrystalline GNRs with different symmetric tilt grain boundaries was investigated in [124] using both NEMD and BTE. Authors demonstrated that thermal conductivity is determined by phonon scattering on edge roughness and grain boundaries. The strong length and temperature dependence of TC was also revealed. The RT thermal conductivity $\sim 4000$ W/mK was found for SLG sheet with 10 - µm length. The substantially lower values of RT TC $\sim 168$ and 277 $Wm^{-1}K^{-1}$ were predicted for pristine and bicrystalline GNRs with the width of 4.1 nm due to the strong phonon scattering on edges and grain boundaries.

Shen et al. [125] carried out theoretical investigation of thermal transport in GNRs, using BTE approach within RTA and Klemens-like formula [103-104] for three-phonon Umklapp scattering. The dependence of phonon frequencies on $q$ was considered as linear for in-plane acoustic phonons and quadratic for out-of-plane acoustic phonons over entire of the BZ. To take into account the selection rules for ZA phonons scattering [40-42] authors of Ref. [125] increased their relaxation time by a factor of 3 because of only 4 types of three-phonon processes from 12 are allowed: ZA+ZA↔LA, ZA+ZA↔TA, LA+ZA↔ZA and TA+ZA↔ZA. The obtained results show that thermal conductivity of GNRs strongly depends on edge roughness, flake length and temperature, which is in a good agreement with the previous theoretical studies [26, 33, 40-41, 105, 108]. The RT TC of 1-µm wide and 10-µm long GNR decreases from ~ 4700 Wm$^{-1}$K$^{-1}$ to ~ 2750 Wm$^{-1}$K$^{-1}$ with the change of phonon-edge scattering from purely specular (*p*=1) to purely diffusive (*p*=0). This decrease in TC value is weaker in comparison with that reported by Nika et al. [108]. The difference can be attributed to different formulas of Umklapp scattering rates used: in Ref. [108] the Umklapp scattering rates were calculated beyond Klemen's-like formula, taking into account all possible 2D three-phonon processes allowed by the momentum and conservation laws.

Another theoretical study of the thermal conductivity of suspended and supported GNRs employed continuum approach for the phonon energy spectra calculations, Callaway formalism for thermal conductivity calculations and standard formulas for Normal, Umklapp, point-defects and rough-edge phonon scatterings [126]. Authors found that for narrow GNRs with width < 50 nm in the energy spectra of acoustic phonons appear many confined branches, resulting in dependence of the average phonon group velocities on the phonon energy. The average velocities in GNRs are close to those in SLG only at small energies $\hbar\omega < \sim 10\,\text{meV}$ and smaller than that in SLG for wide energy range 10 – 180 meV [126]. Another important observation of this study was strong-enough dependence of TC on extrinsic parameters of GNRs: point defects, edge quality and sample size. These findings are in line with many previous theoretical and experimental results [6, 20-21, 26, 33, 40-41, 100-101, 105, 108]. The dependence of the thermal conductivity on temperature for 260 nm-long and 45 nm – thick supported GNR, calculated in Ref. [126] is in agreement with the experimental TC values measured for GNR with the same size in Ref. [100]. It is also important to note here that the decreasing of phonon group velocity in GNRs due to phonon confinement found in Ref. [126] is in qualitative agreement with the theoretical predictions made earlier for semiconductor thin films and nanowires [127-130].

The strain effect on the thermal transport in graphene was theoretically investigated by Lindsay et al. [43] using the Boltzmann-Peierls equation. The authors employed the density functional perturbation theory (DFPT) for calculating the harmonic interatomic force constants required for accurate description of the phonon scattering rates. The results revealed strong dependence of TC on temperature and sample size, and relative weak dependence of TC on weak isotropic tensile strain (~1%). These findings are in line with the previous study [131]. At the same time, stronger dependence of TC on small tensile strain in graphene was predicted in Refs. [53], using continuum approach for phonons and BTE-LWA. The discrepancy may be attributed to difference in the phonon dispersion and phonon scattering rates. The latter confirms that accurate description of phonons in graphene materials is requires for capturing main feature of thermal conductivity. This assessment coincides with the conclusions made by Fugallo et al. [132], which reported on the thermal transport of collective phonon excitations in graphene. According to Fugallo et al. [132], BTE RTA approach strongly underestimates the TC in graphene and overestimates the influence of the strain. In contrast, the thermal conductivities calculated from the exact solution of BTE without single mode RTA, resulting in collective phonon excitations, were in a good agreement with the experimental data. Fugallo et al. [132] also found the strong dependence of TC on the graphene flake length. The high value of RT TC ~ 3500 W/mK which is close to the experimental values [4-5] was obtained for very long flake ($L$ ~ 1 mm).

Many recent theoretical results on thermal transport in graphene were obtained using MD simulations [133-140]. Chen and Kumar [133] investigated thermal transport in graphene supported on copper within equilibrium MD simulations and relaxation time approximation. The interaction with Cu substrate was modeled using the Lennard-Jones potential. The authors found that coupling to substrate significantly influences low-energy and low-wave vector part of phonon energy dispersions in supported SLG as compared with suspended SLG. The RT TC decreases with increasing of interaction strength between carbon and copper atoms from ~ 1800 W/mK (suspended SLG) to 1000 W/mK (supported SLG with strong coupling to substrate). The effect of strain and isotopic disorder on thermal transport in suspended SLG were studied by Pereira and Donadio [134] using equilibrium MD simulations. The authors predicted that the thermal conductivity of unstrained SLG is finite and converges with the sample size at finite temperature. However, TC of the strained graphene diverges logarithmically with the sample size when strain exceeds a threshold value of 2 %. The authors concluded that ZA modes are important for obtaining the finite TC in suspended SLG because they contribute the essential scattering channels to limit the thermal conductivity. In unstrained graphene, the population of ZA modes reduces, while their lifetime increases, resulting in divergence of TC. The authors also shown that isotopic effect strongly influences the thermal conductivity: RT TC decreases from ~ 1000 W/mK to ~ 450 W/mK with increasing of $^{13}$C concentration from 0 (pure $^{12}$C graphene) to 50 %. This result is in a good agreement with experimental measurements [98].

Other MD studies [135-140] confirmed that structural defects of the crystal lattice may significantly suppress the thermal conductivity in graphene and change the temperature dependence of TC. Khosravian et al. [135] found that RT TC of graphene flake decreases from 180 W/mK to 80 W/mK with increasing the number of multi-vacancy defects. Fthenakis et al. [136] demonstrated that TC depends sensitively on whether the defects are isolated, form lines or form extended arrangements in haeckelites. The presence of nonhexagonal rings in crystal lattice made the thermal conductivity anisotropic [136]. According to Fthenakis et al. [136], the TC in graphene with defects can be suppressed up to two orders of magnitude depending on temperature and defects type. Yang et al. [140] considered thermal transport in 21.2 nm long and 3.8 nm wide AGNRs with triangular vacancy (TV) defect and concluded that increase of TV size leads to the suppression of thermal conductivity. It was found that presence of TV defect with 25 removed carbon atoms decreases the RT TC by more than 40% from 230 W/mK to 150 W/mK due to phonon-defect scattering.

The thermal conductivity in nitrogen-doped graphene and GNRs was studied in Refs. [137, 139] within reverse NEMD [137] and EMD based on Green-Kubo method [139]. Yang et al. [137] found that thermal conductivity of N-doped GNRs is smaller than that in GNRs without doping and strongly depends on nitrogen atoms distribution. The RT TC ~ 50 W/mK was found for 11 nm – long and 2 nm – wide GNR with rhombus shape doping, which is by a factor of ~ 1.9 smaller than that in GNR without doping [137]. Goharshadi and Mahdizadeh [139] reported on 59.2% decrease of RT TC in nitrogen-doped graphene with low concentration of $N$ ~ 1%.

Feng et al. [138] carried out theoretical study of the phonon relaxation time, phonon mean-free path and thermal conductivity in defected graphene within the normal mode analysis based on equilibrium MD. Four types of defects were considered: isotopes, Stone-Thrower-Wales (STW) defects, mono vacancies (MV) and double vacancies (DV). The authors have shown that the thermal conductivity strongly decreases in defected graphene: 1.1% of STW (MV) defects suppresses the thermal conductivity by ~ 90% (95%). These findings are in a general accordance with earlier MD simulations [141-143]. The analysis of the frequency dependence of the phonon relaxation time for point-defect scattering revealed deviation from traditionally used dependence: $\tau_{p\text{-}d} \sim \omega^4$ for three-dimensional materials and $\tau_{p\text{-}d} \sim \omega^3$ for two-dimensional materials. According to Feng et al. [138], $\tau_{p\text{-}d} \sim \omega^n$, where *n* depends on the type of defects: $n = 1$ for STW defects and $n=1.1 – 1.3$ for MV and DV defects, with exception of a few long-wavelength phonons, demonstrating $\sim \omega^4$ dependence. The data scatter in the thermal conductivity values in graphene with defects shows that additional investigations are required, because both thermal conductivities and predicted $\tau_{p\text{-}d}(\omega)$ dependence may be strongly affected by the simulation domain size, used in the MD calculations. Wei et al. [144] studied phonon thermal transport in SLG, using spectra-based MD simulations and Tersoff potential for carbon-carbon interaction. Depending on the temperature, TC values in the range 2000 – 8000 W/mK for classical statistics (CS) and 2000 – 3500 W/mK for Bose-Einstein statistics (BES) were obtained (see figure 13). The difference between TC values calculated using BES and CS is large enough at RT and decreases with temperature. Thus, different phonon statistics employed in different MD models may lead to large discrepancy in TC values near RT.

<Figure 13: K vs T >

The electron contribution to heat conduction in graphene has not been studied in details. The first estimates from the experimental data using the Wiedemann – Franz low revealed negligible contribution of electrons to the thermal transport as compared to phonons [4]. Recent calculations within the density functional theory and many-body perturbation theory demonstrated higher values of the electronic thermal conductivity $\kappa_{el}$ at RT [145]: $\kappa_{el} \sim 300$ W/mK, which constitutes around 10% of the total TC. It was also found that electronic TC strongly decreases with the rise of impurity concentration [145]. One should note that strong electrostatic bias of graphene resulting in high concentration of electrons can change the relative contribution of electrons to heat conduction. The theoretical data on thermal conductivity in graphene and GNRs are presented in Table 2 for RT (unless another temperature is indicated).

*II.3. Contribution of different phonon branches to thermal conductivity*

The relative contribution of different phonon polarization branches to thermal conductivity in graphene is one of the most interesting questions related to the physics of thermal transport in 2D crystal lattices. A number of theoretical studies [26, 33, 40-43, 103-105, 125, 126, 133, 138, 144, 159, 161, 162] investigated the polarization branch-dependent thermal conductivity in graphene and GNRs within both BTE approach and MD approaches. Klemens [103-104] assumed that ZA modes carry negligible amount of heat owing their small group velocities and large Gruneisen parameter. The calculations of TC, employing BTE-LWA approach [26], confirmed Klemens' assumption: the contribution of ZA modes to TC was smaller than 1% due to both small group velocities and large values of $|\gamma_{ZA}(q)|$ for long and medium wavelength ZA phonons. However, calculations of TC within BTE-IFC approach predicted dominant role of ZA modes in thermal transport [40-42]. Lindsay et al. [40] found that the contribution of ZA phonons to RT TC in SLG constitutes 75% for 10 µm-long graphene flake and decreases down to 40% in FLG with number of layers $n>4$ [41]. The breaking of SLG selection rules for ZA phonons scattering in FLG was indicated as the primary reason for smaller contribution of ZA phonons to TC [41]. Singh et al. [42] also employed BTE-IFC approach for the investigation of thermal transport in SLG and FLG. Their results [42] were in a good agreement with those obtained by Lindsay et al. [40-41]. At the same time, more recent studies of thermal conductivity in graphene and GNRs reported various values for the relative contributions of ZA modes to TC: from several to 40 %. Moreover, strong dependence of ZA modes contribution to TC on temperature, sample size and defects was revealed.

Lui et al. [124] demonstrated that relative contribution of different phonon branches to TC for bicrystalline GNRs depend on the temperature: at low temperatures ZA modes dominate the thermal transport, while the contribution of LA and TA modes become more important for $T>150$ K. Bae et al. [100] calculated the contribution of different phonon branches to TC in GNRs within the BTE approach. For all considered GNRs, the contribution of ZA phonons to TC was smaller than that of LA or TA modes. The authors also concluded that large intrinsic MFP of LA and TA modes makes them more sensitive to GNR edge disorder while ZA modes are predominantly affected by substrate scattering [100].

Shen et al. [125] found that the contribution of ZA phonons to TC of GNRs varies strongly with temperature and the sample size. In 1-µm wide and 10-µm long GNR, the contribution of the out-of-plane phonons decreases fast with increasing temperature: from 80% at ~10 K to 20 % at 80 K. At temperature $T>100$ K, the in-plane phonons carry 90% of heat. The contribution of the in-plane phonons to thermal conductivity also increases with increasing ribbon length. For 100-µm long GNR, the contribution of TA, LA and ZA phonons to RT TC is ~ 5000, ~3000 and ~ 500 Wm$^{-1}$K$^{-1}$, correspondingly. Thus, in large graphene flakes, the ZA phonons carry less than 5% of heat.

Nissimagoudar and Sankeshwar [126] also concluded that the contribution of different phonon branches to TC of GNRs strongly depends on temperature: ZA modes are the main heat carries for low temperatures $T \leq T_{lim}$, while LA and TA phonons dominate thermal transport for $T>T_{lim}$. The value of $T_{lim}$ depends on the GNR size and is different for suspended and supported GNRs. For 1 µm-long and 5 nm – thick GNRs $T_{lim}$ is ~ 250 K for suspended GNRs and ~150 K for supported GNRs [126]. The deviation from the quadratic phonon dispersion $\omega \sim q^2$ of ZA branch in graphene [38], resulting in lower values of group velocities, could slightly decreasing the relative contribution from ZA phonons to TC [125-126].

MD simulations were also intensively employed for investigating the relative importance of each phonon polarization branch to thermal transport [133, 138, 144, 158, 161-162]. Ong and Pop concluded [158] that coupling to the substrate reduces the thermal conductivity of SiO$_2$-supported graphene by an order of magnitude in comparison with suspended SLG due to the damping of the ZA phonons [158]. This conclusion is in line with findings from Refs. [41-42]. However much smaller contribution from ZA phonons to TC was reported in Refs. [133, 138, 144]. Chen and Kumar [133] calculated that LA, TA and ZA phonons carry ~ 40 %, 20 % and 22 % of total heat, respectively in suspended SLG at RT. The contribution of optical phonons was found as large as ~ 18 %. It was established that in supported SLG the contribution to RT TC changes and constitutes ~ 50% for LA phonons, ~ 21% for TA phonons, ~ 7% for ZA phonons and ~ 22% for optic phonons.

Wei et al. [144] employed the spectral-based MD simulations for analysis of the contributions of different phonon modes to thermal conductivity. It was found that coupling to a substrate reduces the contribution of ZA phonons from 41.1 % in suspended SLG to ~ 20% in supported SLG [144]. Feng et al. [138] investigated the branch –dependent thermal conductivity in defected graphene, using the normal mode analysis based on the equilibrium MD. It was concluded that in pristine SLG, the LA/TA/ZA phonons carry ~35 %/ ~27%/ ~30% of heat, while the contribution from ZO phonons is about 7% [138]. In STW-defected and MV-defected graphene the contribution of ZA modes reduces to 20 %, while contribution from LA modes increases to 50%. The decrease of the contribution of ZA modes in defected graphene was explained by the breakdown of reflection symmetry in the direction perpendicular to graphene layer [138]. At the same time, Gill-Comeau and Levis [161-162], considering the collective phonon excitations in graphene, concluded that ZA phonons dominate the thermal transport, carrying ~ 78 % of heat. Figure 14 illustrate how sensitive the relative contribution of each phonon polarization branch can be to the amount and nature of the defects in graphene.

<Figure 14>

From the review of these theoretical and computational results we can conclude that the studies, employing BTE-LWA approach, usually, predict small contribution from ZA phonons to RT TC due to their small group velocities and overestimation of their scattering. The overestimation comes from two reasons: (1) omission of ZA selection rules and (2) large values of $|\gamma_{ZA}(q)|$ for long- and medium wavelength phonons $\gamma_{ZA}^2(q) >> \gamma_{LA}^2(q), \gamma_{TA}^2(q)$ [57], resulting in shorter life-time for ZA phonons $\tau_U(q) \sim 1/\gamma^2(q)$. In contrast, BTE-IFC studies predict the dominant role of ZA phonons in thermal transport in suspended graphene owing their weaker scattering as compared with LA and TA modes. Weak scattering of ZA phonons is explained both by the ZA selection rule, which limits scattering, and small values of 3-rd order IFCs for a certain modes. The high – order unharmonic processes, which are not taken into consideration in BTE-LWA and BTE-IFC, could significantly change the relative branch contribution. The latter is confirmed by several independend MD studies, predicting larger contribution to TC from the in-plane acoustic phonons [133,138,144]. However, other MD studies demonstrate opposite view [158,161-162]. Including collective excitations in the thermal transport models significantly change the final concluions [132,161-162]. Another reasons for scatter of thermlal conductivity data in diffrent models are (i) various interatomic potentials used, (ii) difference in the simulation domain size in MD approaches and (iii) different formulation of the heat auto-correlation functions. The discrepancy in reviewed results suggests that additional theoretical and experimental studies are required to shed light on relative contributions from different branches to thermal conductivity of graphene materials. In Table 3 we summarize the branch-dependent contributions to thermal conductivity reported up to date. The data are presented for the room temperature unless different temperature is indicated.

*Acknowledgements*

This was supported as part of the Spins and Heat in Nanoscale Electronic Systems (SHINES), an Energy Frontier Research Center funded by the U.S. Department of Energy, Office of Science, Basic Energy Sciences (BES) under Award # SC0012670. Authors thank Dr. A. Cocemasov for technical help with Figure 4.

**Table 1.** Thermal conductivity of graphene and graphene-based materials: experimental data

| $\kappa$ (W/mK) | Method | Brief description | Ref. |
|---|---|---|---|
| *Single-layer graphene* | | | |
| ~3000 – 5000 | Raman optothermal | suspended; exfoliated | 4,5 |
| 2500 | Raman optothermal | suspended; chemical vapor deposition (CVD) grown | 90 |
| 1500 – 5000 | Raman optothermal | suspended; CVD grown | 91 |
| 600 | Raman optothermal | suspended; exfoliated; $T \sim 660$ K | 92 |
| 2000 – 3800 | electrical self-heating | exfoliated and CVD grown; $T\sim300$ K | 93 |
| 310 + 200/-100 | | exfoliated and CVD grown; $T\sim1000$ K | |
| 600 | electrical | supported; exfoliated | 94 |
| 1600 – 2800 | Raman optothermal | suspended; strong isotope dependence; $T\sim380$ K | 98 |
| 2778.3 ± 569 | Raman optothermal | suspended, $T \sim 325$ K | 99 |
| *Few-layer graphene* | | | |
| 1300 – 2800 | Raman optothermal | suspended FLG; exfoliated; $n$=2–4 | 34 |
| 50 – 970 | heat-spreader method | FLG, encased within $SiO_2$; n = 2, …, 21; $T\sim310$ | 95 |
| 560 – 620 | electrical self-heating | suspended bilayer graphene; polymeric residues on the surface | 96 |
| 302 – 596 | modified T-bridge | suspended FLG; $n$=2 – 8 | 97 |
| 1896 ± 390 | Raman optothermal | suspended bilayer graphene; $T \sim 325$ K | 99 |
| 1412.8 ± 390 | | suspended twisted bilayer graphene; $T \sim 325$ K | |
| (730 – 880) ± 60 | electro-thermal micro-bridge method | suspended bilayer graphene; polymeric residues on the surface; 13 µm long and 5 µm thick | 109 |
| 150 – 1200 | electrical self-heating | suspended and supported FLG; polymeric residues on the surface | 115 |
| *Graphene nanoribbons* | | | |
| 80 – 230 | electrical self-heating | supported; strong size dependence | 100 |
| ~1500 | electro-thermal micro-bridge method | suspended, CVD grown; 9 µm long; logarithmic dependence on the sample length | 101 |
| 205 – 2236 | electrical four-wire method | TC increases with sample width decrease | 102 |
| 1100 | electrical self-heating | supported; exfoliated; $n$<5 | 116 |
| 80 – 150 | electrical self-heating | $SiO_2$ – supported; dependence on the edge roughness and defects | 117 |

**Table 2.** Thermal conductivity in graphene and graphene nanoribbons: theoretical models

| κ (W/mK) | Method | Brief description | Refs |
|---|---|---|---|
| *Single-layer graphene* | | | |
| 2000-8000 | BTE–LWA + all possible three-phonon transitions | TC dependence on edge roughness, flake width and Grunaisen parameter | 26 |
| 1500 – 3500 | BTE–IFC | TC dependence on flake size | 40 |
| ~ 3100 | BTE–IFC + density function perturbation theory | $L = 10$ μm; TC dependence on flake size; weak dependence of TC on small isotopic strain (<1%) | 43 |
| 2000 – 4000 | BTE–LWA + continuum approach | strong isotope, point-defects and strain influence | 52,53 |
| 4400 | BTE–LWA | average $\gamma = 4$ and average $<\upsilon> = 18.6$ km/s for in-plane phonons; strong size dependence $\kappa$~log(L) | 103,104 |
| 1000 – 8000 | BTE–LWA | different average $\gamma$ and different group velocities for in-plane phonons; strong size dependence $\kappa$~log(L) | 105 |
| 100 – 8000 | BTE–IFC | TC dependence on flake size, shape and edge roughness | 108 |
| 1000 | EMD | Brenner-type bond order interatomic potential (IP) | 118 |
| ~ 1500 | MD | Tersoff-Brenner potential for C-C IP | 119 |
| ~ 6600 | EMD and NEMD | Tersoff IP | 120 |
| 4000 – 6000 | BTE-LWA | strong dependence on strain larger than 4% | 131 |
| ~ 3500 | BTE beyond RTA; collective phonon excitations | flake length ~ 1 mm; strong length dependence; weak dependence on strain and weak dependence on length in strained graphene | 132 |
| 1800 | EMD | 6 nm × 6 nm sheet; isolated | 133 |
| 1000 – 1300 | EMD | 6 nm × 6 nm sheet; Cu – supported; strong dependence on the interaction strength between graphene and substrate | 133 |
| ~1000 | EMD | strong isotopic effect | 134 |
| 300 – 500 | NEMD | strong defect influence | 136 |
| 2900 | NEMD | strong dependence on the vacancy concentration | 141 |
| ~3300 | spectra-based MD + Tersoff IP | Bose-Einstein statistics | 144 |
| ~8000 | | classical statistics | 144 |
| ~2430 | BTE-IFC | $\kappa(graphene) \geq \kappa(carbon\,nanotube)$ | 146 |
| 4000 | ballistic | strong width dependence | 147 |
| 20000 | VFF + ballistic regime | flake length ~ 5 µm; strong width and length dependence | 148 |
| 100 – 550 | NEMD | flake length $L$<200 nm; strong length and defect dependence | 149 |
| 3000 | NEMD | flake length ~ 15 µm; strong size dependence | 150 |
| 2360 | NEMD | $L$~5 µm; strong length dependence | 151 |
| 100 – 600 | non-equilibrium Green functions | strong dependence on grain size and line defects | 152 |
| ~ 256 | EMD and NEMD | TC in $SiO_2$–supported SLG is by an order of magnitude lower than in suspended SLG | 158 |
| electronic TC ~ 300 | density functional theory + many-body perturbation theory | strong dependence on the impurity | 145 |

*Few-layer graphene*

| | | | |
|---|---|---|---|
| 1000 – 4000 | BTE-LWA, $\gamma_s(q)$ | n = 8 – 1, strong size dependence | 34 |
| 1000 – 3500 | BTE-IFC | n = 5 – 1, strong size dependence | 41 |
| 2000 – 3300 | BTE-IFC | n = 4 – 1 | 42 |
| 580 – 880 | NEMD | n = 5 – 1, strong dependence on the Van-der Vaals bond strength | 157 |

*Graphene nanoribbons*

| | | | |
|---|---|---|---|
| 5500 | BTE-LWA | GNR with width of 5 µm; strong dependence on the edge roughness | 33 |
| 400 – 600 | NEMD | $K$~$L^{0.24}$; 100 nm ≤ $L$ ≤650 nm | 111 |
| 2000 | MD + Brenner IP | $T$=400 K; 1.5 nm × 5.7 nm zigzag GNR; strong edge chirality influence | 121 |
| 200 – 900 | NEMD + Tersoff IP | strong strain and length dependence $\kappa$~$L^n$, where $n = 0.47$ for 20-AGNR and $n = 0.35$ for 10-ZGNR; $L$=10 nm – 60 nm | 122 |
| 168 – 4000 | BTE-LWA, NEMD | bicrystalline GNRs; 4.1 nm ≤ $L$ ≤ 10 µm TC dependence on length | 124 |
| 2750 – 4000 | BTE-LWA | GNRs with $L$ = 10 µm and W = 1 µm; dependence on edge roughness | 125 |
| 60 – 70 | BTE-LWA | narrow GNRs with W<50 nm; confined phonon branches; strong edge scattering | 126 |
| ~ 50 | reverse NEMD | nitrogen-doped 11 nm – long and 2 nm – wide GNRs; strong dependence on nitrogen atoms distribution | 137 |
| ~ 2500 | EMD + Tersoff IP | nitrogen-doped 10 nm – long and 2 nm – wide GNRs; strong dependence on nitrogen concentration | 139 |
| ~ 230 | reverse NEMD | 21.2 nm – long and 3.8 nm – wide AGNRs; strong dependence on triangular vacancy size | 140 |
| 1000 – 7000 | EMD + Tersoff IP | strong ribbon width and edge dependence | 153 |
| 30 – 80 | reverse NEMD + AIREBO IP | 10 - zigzag and 19 -arm-chair nanoribbons; strong defect dependence | 154,155 |
| 3200 – 5200 | EMD | strong GNRs width ($W$) and length dependence; 9 nm $\leq L \leq 27$ nm and 4 nm $\leq W \leq 18$ nm | 156 |
| 100 – 1000 | BTE-LWA | $SiO_2$ – supported GNRs; strong edge and width dependence | 159 |
| 500 – 300 | NEMD | few-layer G10-ZGNR, $n = 1,...,5$ | 160 |

**Table 3.** Contribution of different phonon polarization branches to thermal conductivity in graphene and GNRs

|  | **Model** | **Description** | **Ref.** |
|---|---|---|---|
| SLG | BTE–LWA | Assumption that only LA and TA phonons participate in thermal transport (TT) | 103, 104, 105 |
| SLG | BTE–LWA | LA + TA ~ 99 % | 26 |
| SLG | BTE–IFC | ZA ~ 75%; TA ~ 15%; LA ~ 9 % | 40 |
| FLG | BTE-IFC | ZA contribution decreases from ~ 75 % for SLG to 38 % for 6-layer FLG; TA and LA contribution is insensitive to number of layers $n$: TA ~ 15 % and LA ~ 9% | 41 |
| FLG | BTE-IFC | ZA phonons are dominate heat carries; thermal conductivity decreases with rise of $n$ owing decrease of ZA phonon contribution | 42 |
| SLG | BTE-LWA | strong dependence of ZA contribution on the temperature; ZA phonons are the main heat carriers for T<50 K, while for T>200 K in-plane acoustic phonons dominate TT; at RT TA/LA/ZA contribution is ~ 65 % / 25 % / 10 % | 33 |
| SiO$_2$ – sipported GNRs | BTE-LWA | strong dependence of ZA contribution on the temperature; ZA phonons dominate thermal transport for T < 100 K, while for T > 200 K TA and LA phonons are the main heat carriers | 159 |
| SLG | BTE-IFC | flake length $L = 10$ µm; ZA ~ 76 %, LA + TA ~ 20 % | 43 |
| GNRs | BTE | ZA contribution is smaller than TA or LA contribution | 100 |
| GNRs | BTE-LWA | strong temperature and flake size dependence; ZA contribution < ~ 5% for large flakes | 125 |
| GNRs | BTE-LWA | strong dependence of ZA contribution on the temperature and flake size; ZA modes dominate TT at low temperatures, while TA and LA modes are dominant heat carriers for medium and high T. | 126 |
| suspended and SiO$_2$-supported graphene | EMD and NEMD | ZA phonons dominate TT | 158 |
| SLG | EMD | LA ~ 40 %, TA ~ 20 %, ZA ~ 22%, optic phonons ~ 18 % in suspended SLG; LA ~ 50 %, TA ~ 21 %, ZA ~ 7 %, optic phonons ~ 22 % in supported SLG | 133 |
| SLG | spectra-based MD | ZA ~ 41.8 % in suspended SLG and ~ 20 % in supported SLG | 144 |
| SLG | EMD | LA ~ 35 %, TA ~ 27 %, ZA ~ 30 %, ZO ~ 7 % in pristine SLG; | 138 |

| | | LA ~ 50 %, ZA ~ 20 % in SLG with defects | |
|---|---|---|---|
| SLG | EMD + time – domain TC | collective phonon excitations; ZA contribution ~ 78 % | 161,162 |

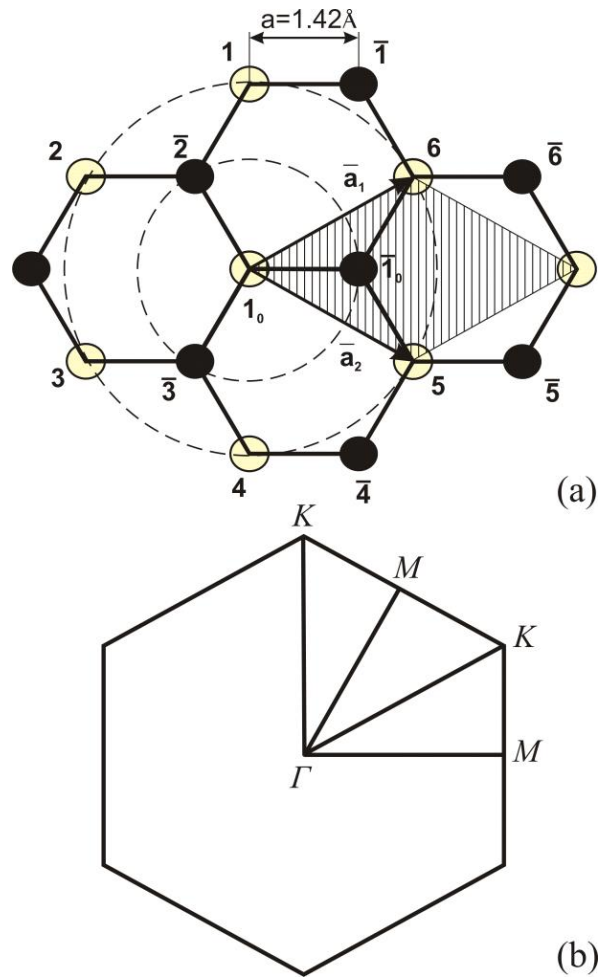

**Fig. 1**. Schematic view of crystal lattice (a) and Brillouin zone (b) for single layer graphene. The figure is reprinted from Ref. [26] with permission from the America Physical Society.

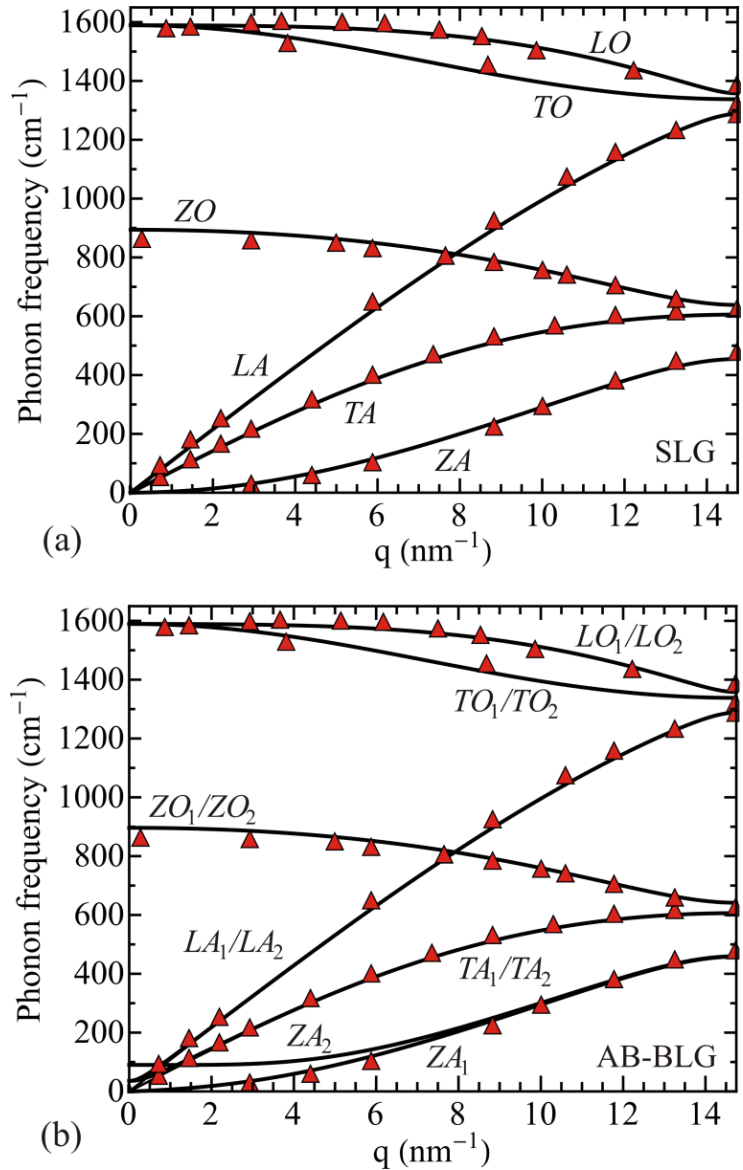

**Fig. 2.** Phonon dispersions in (a) single layer graphene and (b) AB-stacked bilayer graphene, plotted along ΓK direction of Brillouin zone. The phonon energies were calculated using the BvK model of the lattice vibrations. The figure is adopted from Ref. [38] with permission from the Royal Society of Chemistry.

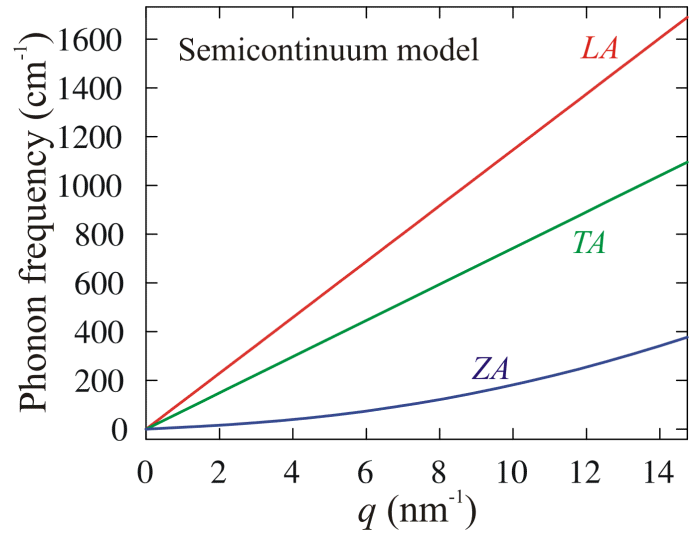

**Fig. 3**. Phonon dispersion in single layer graphene, calculated in the framework of the elastic continuum approach.

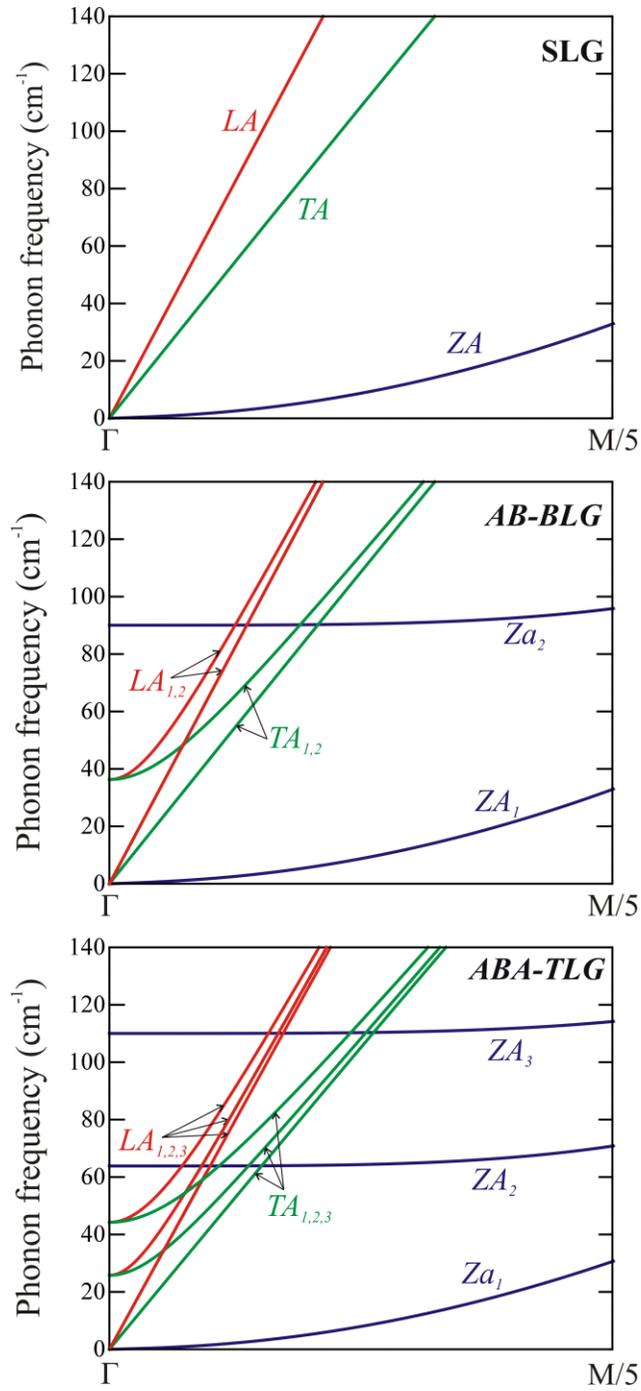

**Fig. 4**. Phonon dispersion in single layer graphene, AB-bilayer graphene and ABA-three-layer graphene, plotted along ΓK direction near the center of Brillouin zone. The phonon energies were calculated using the BvK model of the lattice vibrations.

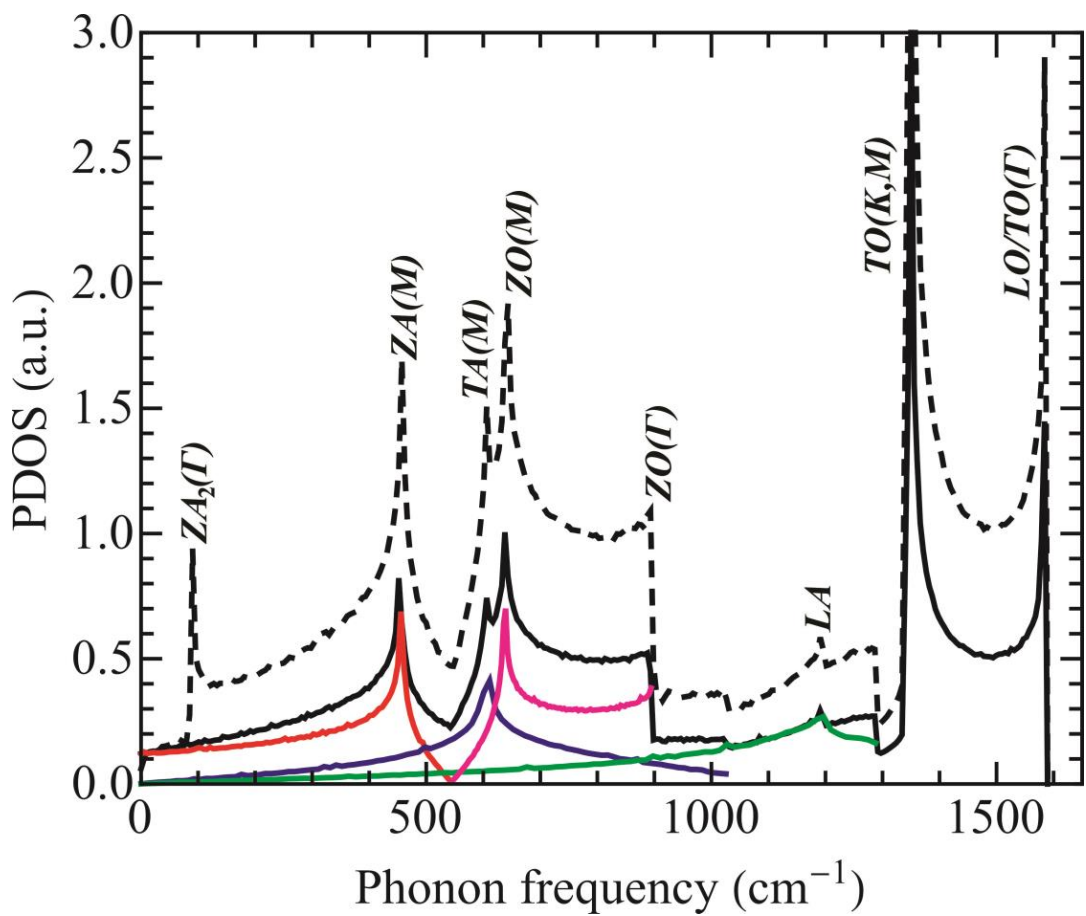

**Fig. 5.** Total phonon density of states in SLG (solid black) and AB-BLG (dashed black), and contribution from ZA (red), TA (blue), ZO (magenta) and LA (green) phonon branches. PDOS is calculated using the phonon dispersion obtained within the BvK model of the lattice vibrations. The figure is reprinted from Ref. [38] with permission from the Royal Society of Chemistry.

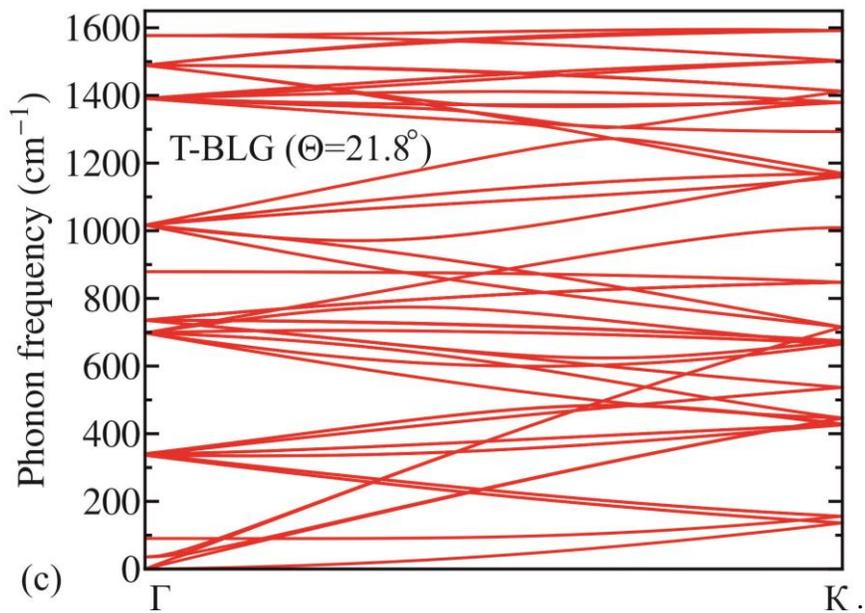

**Fig. 6**. Phonon dispersion in twisted bilayer graphene with the twisting angle 21.8°. The figure is adopted from Ref. [37] with permission from the American Institute of Physics.

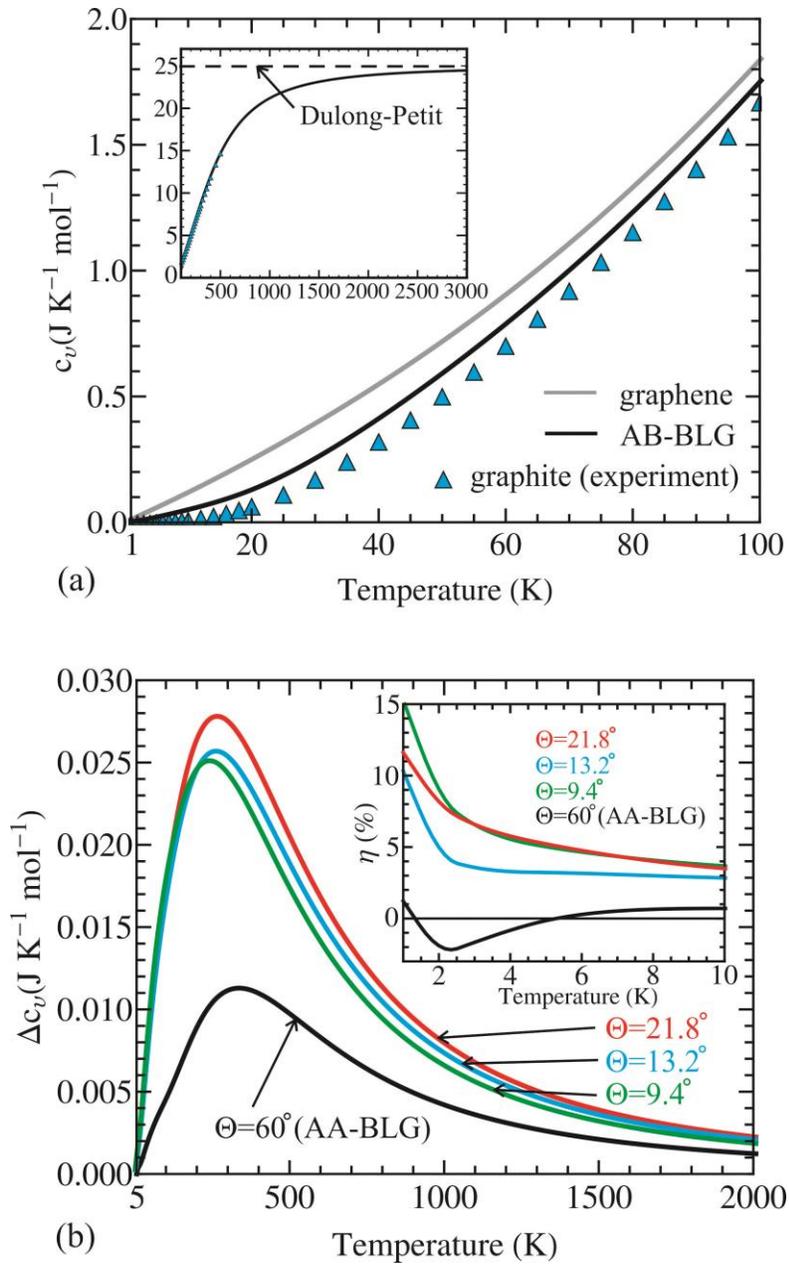

**Fig. 7.** Temperature dependence of **the** phonon specific heat in graphite, SLG, AB-BLG and T-BLG. The figure is adopted from Ref. [37] with permission from the American Institute of Physics.

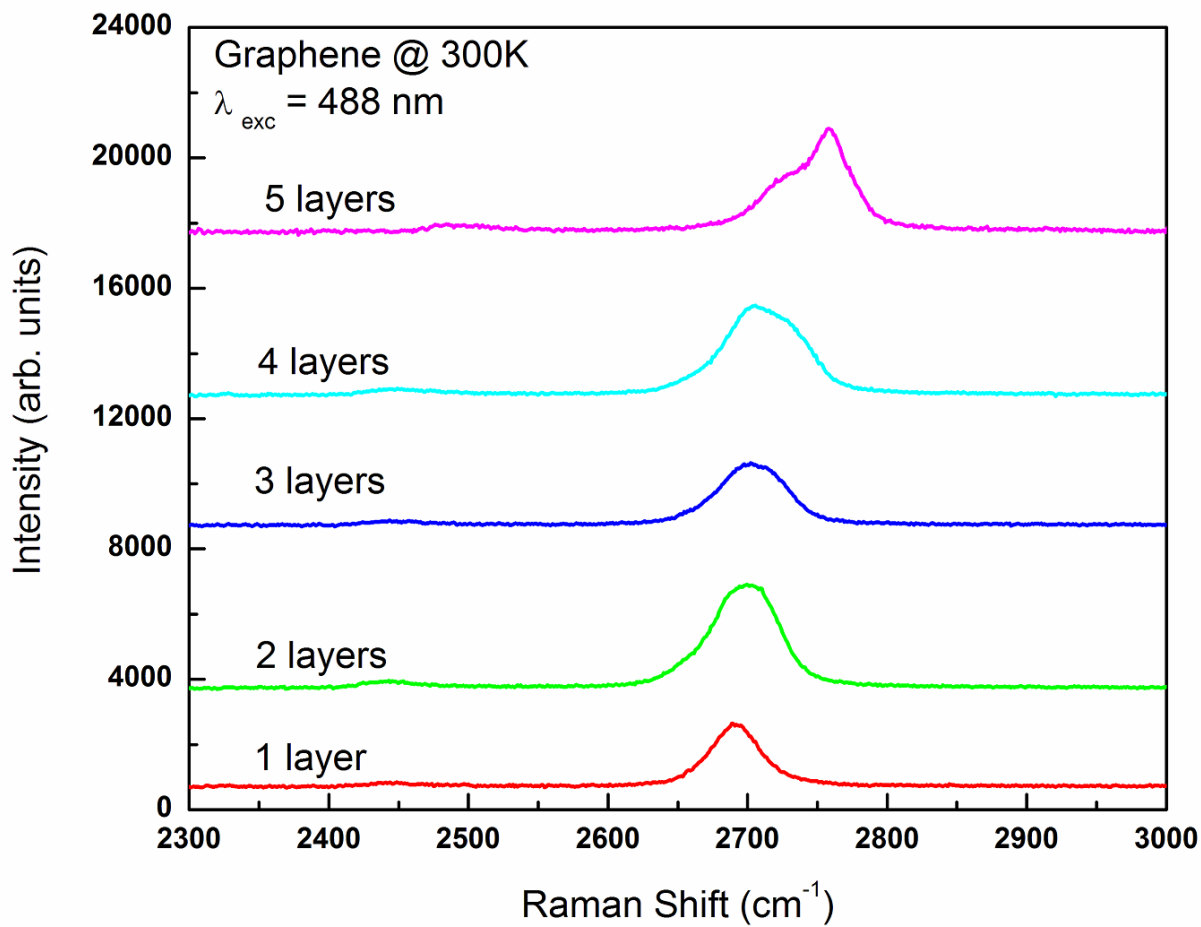

**Fig. 8.** Room-temperature Raman spectrum of graphene and few-layer graphene showing the 2D-band spectral region. Note that the position and shape of the 2D peak depend on the number of atomic planes. The figure is reprinted from Ref. [74] with permission from the American Chemical Society.

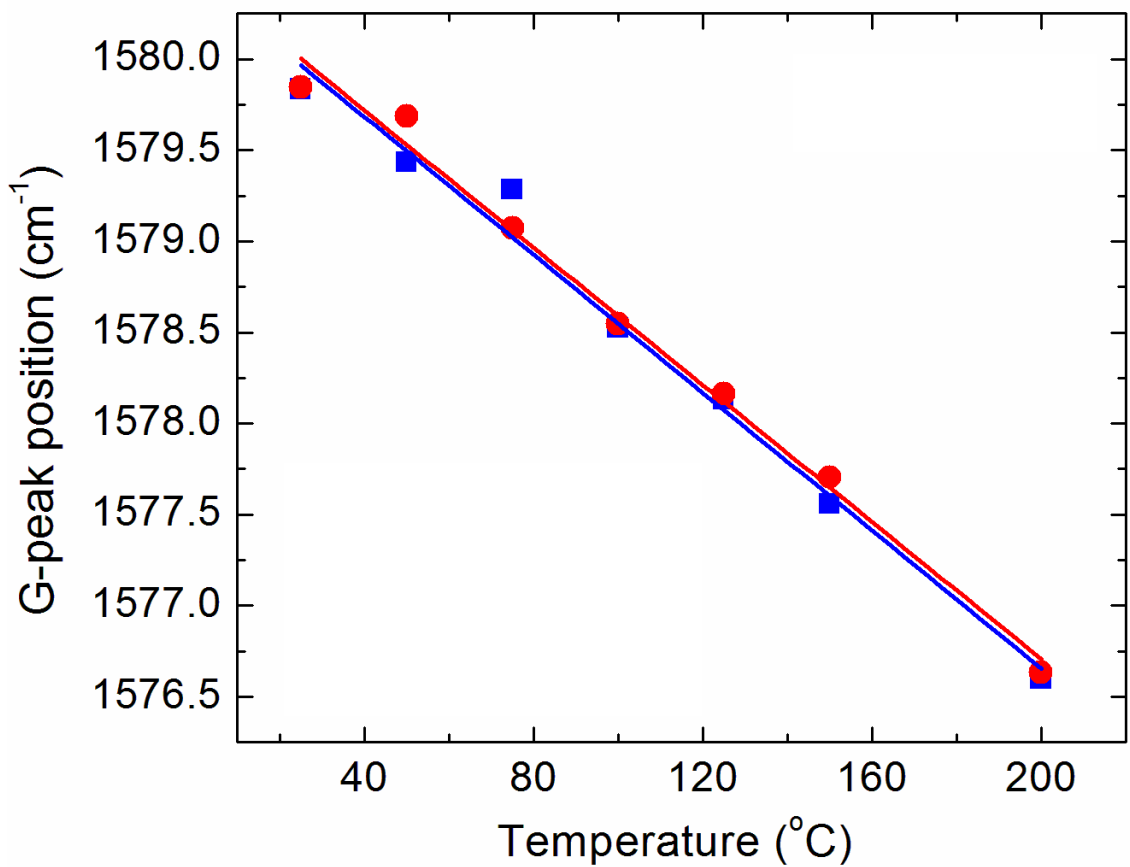

**Fig. 9.** Raman G peak as a function of the sample temperature. The measurements were carried out under the low excitation power to avoid local heating while the temperature of the sample was controlled externally. Note an excellent liner fit for the examined temperature range. The figure is reprinted from Ref. [112] with permission from the American Chemical Society.

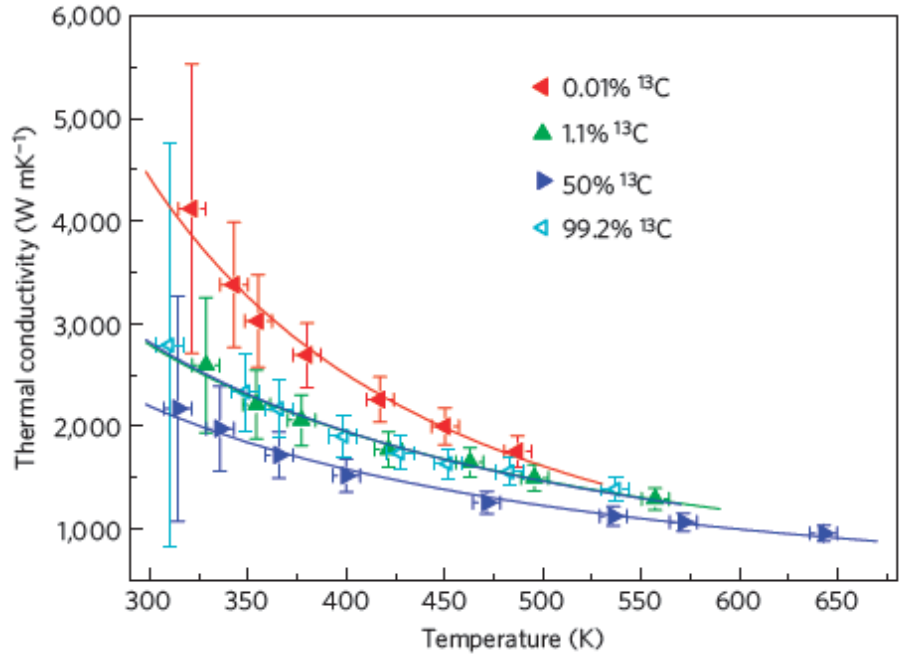

**Fig 10.** Thermal conductivity of suspended graphene with different concentration of $^{13}$C isotope. The figure is reprinted from Ref. [98] with permission from the Nature Publishing Group.

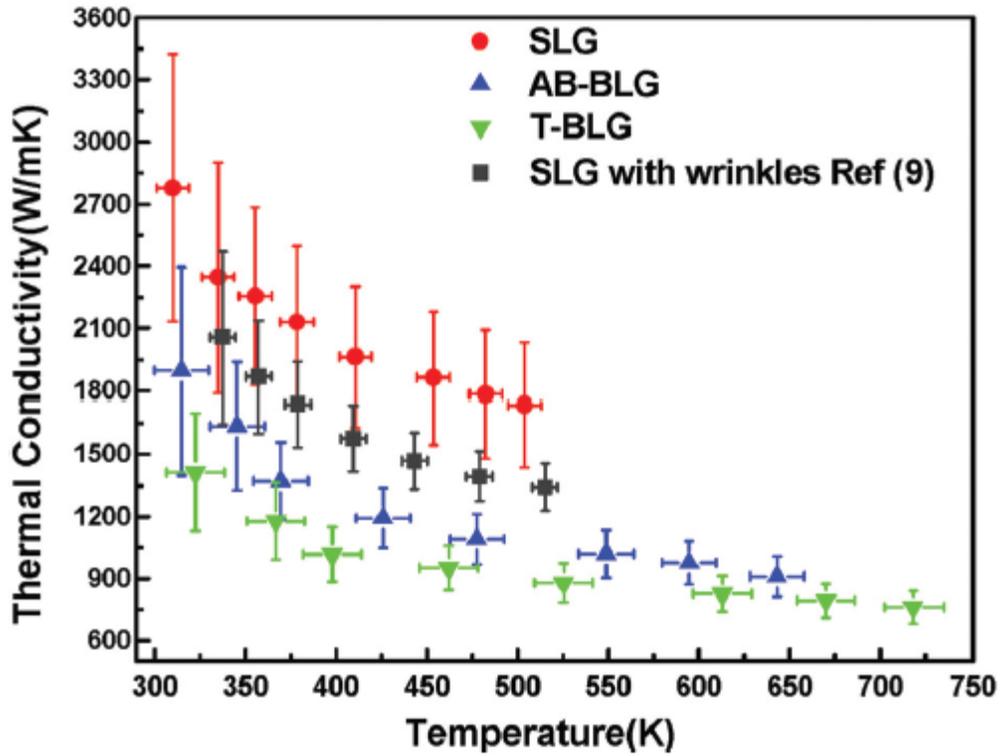

**Fig. 11**. Thermal conductivity of suspended single-layer graphene, AB-bilayer graphene and twisted bilayer graphene as a function of temperature. The figure is reproduced from Ref. [99] with permission from the Royal Society of Chemistry.

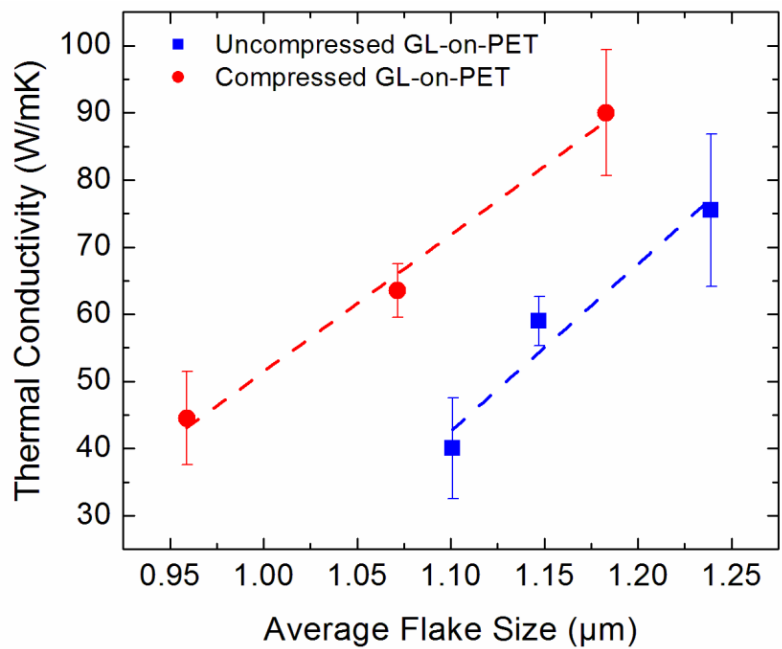

(a)

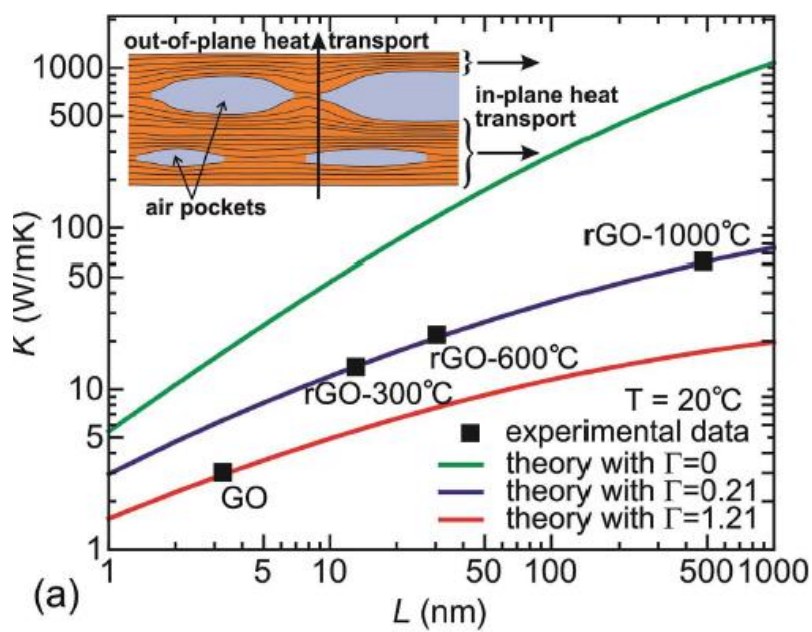

(b)

**Fig. 12**. In-plane thermal conductivity in (a) graphene laminate and (b) reduced graphene oxide as a function of the graphene flake (domain) size. The figures are adopted from Refs. [112-113] with permissions from the American Chemical Society and Wiley VCH Verlag GmbH & Co.

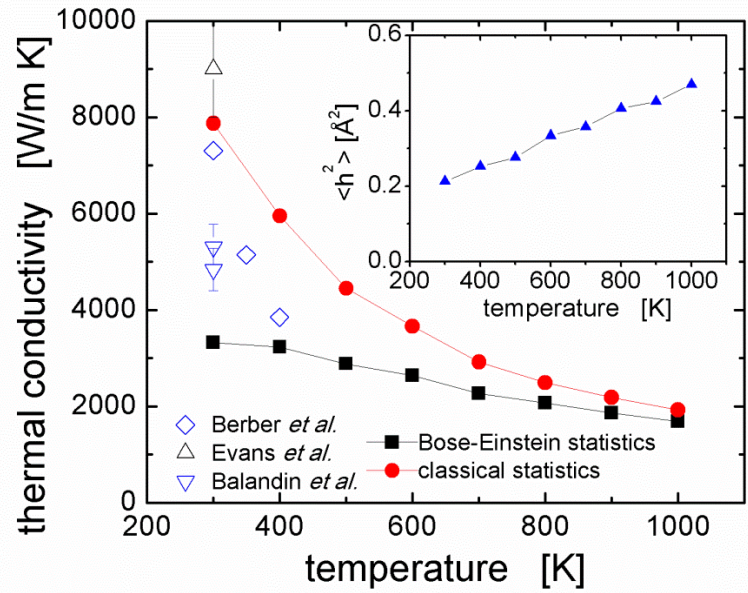

**Fig. 13.** Thermal conductivity of suspended single-layer graphene as a function of temperature calculated for the phonon number obeying the Bose-Einstein and classical statistics. The insert shows the temperature dependent out-of-plane displacement. The data points from Refs. [4, 120, 153] are shown by diamonds and triangles. The figure is reproduced from Ref. [144] with the permission from the American Institute of Physics.

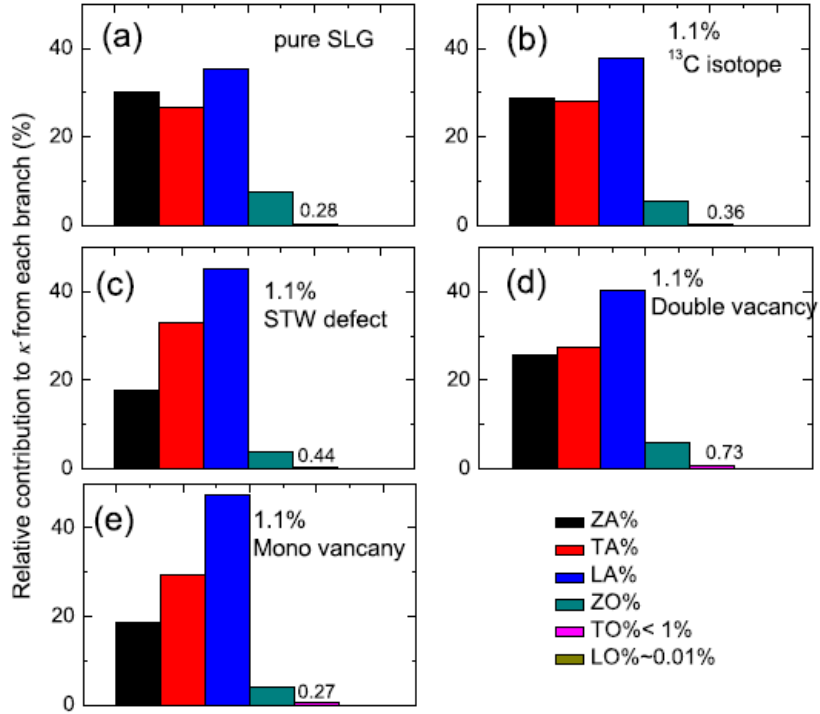

**Fig. 14.** Relative contribution of the phonon branches to the thermal conductivity of graphene calculated using EMD. Note that various theoretical approaches give a wide range of relative contributions depending on the assumptions used. External factors such as presence or absence of a substrate and nature of the defects also affect the relative contributions. The figure is reproduced from Ref. [43] with the permission from the American Physical Society.